# Symmetry-Induced Universal Momentum-Transfer Dependencies for Inelastic Neutron Scattering on Anisotropic Spin Clusters

Shadan Ghassemi Tabrizi

*Technische Universität Berlin, Institut für Chemie, Theoretische Chemie, Sekr. C7, Strasse des 17. Juni 135, 10623 Berlin, Germany*

**Abstract.** Inelastic neutron scattering (INS) is a key method for studying magnetic excitations in spin systems, including molecular spin clusters. The method has significantly advanced in recent years and now permits to probe the scattering intensity as a function of the energy transfer and the momentum-transfer vector **Q**. It was recently shown that high molecular symmetry facilitates the analysis of spectra. Point-group symmetry imposes selection rules in isotropic as well as anisotropic spin models. Furthermore, the **Q**-dependence of the INS intensity may be completely determined by the point-group symmetry of the states involved in a transition, thereby affording a clear separation of dynamics (energies, transition strengths) and geometrical features (**Q**-dependencies). The present work addresses this issue for anisotropic spin models. We identify a number of cases where the **Q**-dependence is completely fixed by the point-group symmetry. For six- and eight-membered planar spin rings and two polyhedra (the cube and the icosahedron) we tabulate and plot the corresponding powder-averaged universal intensity functions. The outlined formalism straightforwardly applies to other highly-symmetric systems and should be useful for future analyses of INS spectra by focusing on those features that contain information on either spin dynamics or the point-group symmetry of states.

## 1. Introduction

The interest in exchange-coupled spin clusters has hugely expanded over the last thirty years [1–3], because the intriguing properties of magnetic molecules could become a central ingredient for future technology [4–13]. Inelastic neutron scattering (INS) was established early on as a technique for the determination of exchange interactions in small spin clusters with unrivalled accuracy [14–16]. The increasing sensitivity and resolution that can be achieved with modern spectrometers and detectors has more recently allowed to quantify a larger range of microscopic interactions in terms of spin-Hamiltonian parameters, including local zero-field splitting or anisotropic exchange (see Refs. [17–19] for reviews). With the exception of Ref. [20], experiments were restricted to powder probes, where the scattering intensity can be recorded as a function of the energy transfer (*E*) and the magnitude of momentum transfer (*Q*). Significant recent progress in different fields (not least instrumentation) now allows to conduct experiments on single crystals to measure *E* and the 3D momentum-transfer vector **Q** in what is known as four-dimensional (4D) INS [19]. This technique is now gaining importance as it provides unique insights into the physics of diverse spin clusters [19,21–25]. Although it was argued that the powder-averaged intensity in principle contains all relevant information (with respect to pure magnetic scattering), in practice 4D-INS spectra facilitate the extraction of such information [26].

When molecular spin clusters comprise a large number of centers, extensive spectral fittings based on exact spin-Hamiltonian eigenstates are prohibited, and a search for simplifications in the theoretical modeling is worthwhile. An exploitation of symmetry indeed offers significant potential in this respect (beyond the symmetry-adaptation of basis states to facilitate exact diagonalization [27–30]). First, spin symmetry is approximately valid in many first-row transition-metal complexes [1], leading to $\Delta S = 0, \pm 1$ INS selection rules. Additional selection rules result from point-group symmetry in isotropic models [31], where point-group symmetry manifests as spin-permutational symmetry, SPS [27]; in specific cases, complete permutational symmetry within a subset of sites leads to further restrictions on INS transitions [32–35]. Secondly, it was found that the **Q**-dependence is completely fixed by the "relative" symmetry of the levels involved in a transition in symmetric spin rings [36]. This idea was recently generalized and applied to a larger class of systems [31].

It is necessary to step beyond spin symmetry when explicitly considering magnetic anisotropy, which has become a focus of interest in molecular magnetism and is amenable to different spectroscopic methods, including INS. Anisotropic interactions (whose physical origin is usually spin-orbit coupling) mix spin states and thereby lift spin-selection rules. SPS is broken too, unless one considers a model with artificially high symmetry, e.g., one that conserves $\hat{S}_z$ [27]. However, combinations of global spin rotations and permutations corresponding to the real-space molecular point-group leave the general anisotropic (zero-field) Hamiltonian invariant [37–39]. INS selection rules may result from this rotational-permutational symmetry [31]. The main objective of the present work is to show how **Q**-dependencies can be fixed by symmetry in anisotropic spin clusters, such that specific **Q**-dependent functions describe all transitions of a definite type. We identify universal **Q**-dependencies in spin rings and polyhedra and tabulate and plot powder-averaged intensity functions for planar rings with six or eight centers, and for the cube and the icosahedron. For a transition associated with a universal **Q**-dependent function, information on spin dynamics is exclusively contained in the energy transfer and the transition strength, while the momentum-transfer dependence (for single crystals or powders) contains only information on the relative point-group symmetry species of the two levels.

## 2. Theory

Usually, isotropic exchange of Heisenberg type represents the leading contribution to the spin Hamiltonian of a multinuclear spin cluster [1],

$$\hat{H}^{(0)} = \sum_{i<j} J_{ij} \hat{\mathbf{s}}_i \cdot \hat{\mathbf{s}}_j \ , \tag{1}$$

where symmetry-equivalent pairs have the same coupling constant $J_{ij}$. An anisotropic spin Hamiltonian that is in accord with the molecular point-group can be constructed by considering symmetric anisotropic coupling,

$$\hat{H}^{(1)} = \sum_{i<j} \hat{\mathbf{s}}_i \cdot \mathbf{D}_{ij} \cdot \hat{\mathbf{s}}_j \ . \tag{2}$$

$\mathbf{D}_{ij}$ is a traceless rank-two tensor,

$$\mathbf{D}_{ij} = D_{ij}(\mathbf{n}_{ij}\mathbf{n}_{ij}^T - \tfrac{1}{3}\mathbf{1}) \ , \tag{3}$$

where $\mathbf{n}_{ij} = \mathbf{R}_{ij}/|\mathbf{R}_{ij}|$ points from site $i$ to site $j$ and the coupling constant $D_{ij}$ is the same for equivalent pairs. In several low-symmetry cases, $\hat{H}^{(0)} + \hat{H}^{(1)}$ has an artificially high symmetry (e.g. in certain dimers). Then, local zero-field splitting $\hat{\mathbf{s}}_i \cdot \mathbf{D}_i \cdot \hat{\mathbf{s}}_i$ (for $s_i > \tfrac{1}{2}$) or antisymmetric exchange $\mathbf{d}_{ij} \cdot \hat{\mathbf{s}}_i \times \hat{\mathbf{s}}_j$ is needed to lower the symmetry group of the spin Hamiltonian to become isomorphic to the molecular point group (the respective double group must be considered for systems with an odd electron number [38]). However, for the systems studied in the present work, antisymmetric exchange is either excluded on account of point-group symmetry (cube, icosahedron, cf. the Moriya rules [1,40]), or not needed to avoid an artificially high symmetry of the spin Hamiltonian (planar spin rings). As our focus is on universal **Q**-dependencies, a choice of specific values for $J_{ij}$ or $D_{ij}$ is of no concern.

As mentioned above, spin-symmetry and SPS, which overall lead to a point-group classification of spin multiplets [27,41], are separately broken by general anisotropic interactions, but combinations of SPS operations with appropriate spin rotations remain intact. INS selection rules are identified based on a symmetry classification of eigenstates and transition operators, where the latter are spanned by the set of local spin operators $\{\hat{\mathbf{s}}_i\}$ [31]. Waldmann has addressed universal **Q**-dependencies in spin rings with uniaxial anisotropy (conserving $\hat{S}_z$ and SPS symmetry) [36]. In the following, we shall elucidate how universal **Q**-dependencies arise more generally for anisotropic spin models that break $\hat{S}_z$ and SPS symmetry.

In the differential neutron-scattering cross-section, Eq. (4) [36,42],

$$\frac{d^2\sigma}{d\Omega d\omega} = \frac{\gamma e^2}{m_e c^2}\frac{k'}{k} e^{-2W(\mathbf{Q},T)} \sum_{n,m} \frac{e^{-E_n/kT}}{q(T)} I_{nm}(\mathbf{Q})\delta\left(\omega - \frac{E_m - E_n}{\hbar}\right) , \qquad (4)$$

$\Omega$ denotes the solid angle, $\hbar\omega$ is the energy transfer, $\mathbf{Q} = \mathbf{k} - \mathbf{k}'$ is the scattering vector, $e^{-2W(\mathbf{Q},T)}$ and $e^{-E_n/kT}/q(T)$ are the Debye-Waller factor and the Boltzmann factor, respectively, and all other symbols have their usual meaning. $I_{nm}(\mathbf{Q})$ is defined in Eq. (5),

$$I_{nm}(\mathbf{Q}) = \sum_{i,j} F_i^*(Q) F_j(Q) e^{i\mathbf{Q}\cdot(\mathbf{R}_i - \mathbf{R}_j)} \sum_{\alpha,\beta}\left(\delta_{\alpha\beta} - \frac{Q_\alpha Q_\beta}{Q^2}\right)\langle n|\hat{s}_{i\alpha}|m\rangle\langle m|\hat{s}_{j\beta}|n\rangle , \qquad (5)$$

where $\alpha, \beta = x, y, z$ and $\mathbf{R}_i$ is the position vector of the $i$-th spin center (the sums over $i$ and $j$ run over all $N$ sites), and $|n\rangle$ and $|m\rangle$ are energy eigenstates. In defining the quantity $L_{mn}(\mathbf{Q}) = |F(Q)|^2 I_{nm}(\mathbf{Q})$, we assume a uniform form factor $F(Q)$ for all ions. The focus will be on $L_{mn}(\mathbf{Q})$, which defines the **Q**-dependence (we ignore form factors, which are known with good accuracy for many ion types).

The spherical integral relevant for powder samples,

$$\bar{L}_{nm}(Q) \equiv \int \frac{L_{nm}(\mathbf{Q})}{4\pi} d\Omega , \qquad (6).$$

was solved based on tensor-operator techniques [36] and later translated to a readily usable Cartesian formulation, which for convenience is quoted from the work of Caciuffo et al. [43] in Eq. (7):

$$L_{nm}(Q) = \sum_{i,j} \Big\{ \tfrac{2}{3}\Big[ j_0(QR_{ij}) + C_0^2 j_2(QR_{ij}) \Big] \tilde{s}_{zi}\tilde{s}_{zj}$$
$$+ \tfrac{2}{3}\Big[ j_0(QR_{ij}) - \tfrac{1}{2} C_0^2 j_2(QR_{ij}) \Big] (\tilde{s}_{xi}\tilde{s}_{xj} + \tilde{s}_{yi}\tilde{s}_{yj})$$
$$+ \tfrac{1}{2} j_2(QR_{ij}) \Big[ C_2^2(\tilde{s}_{xi}\tilde{s}_{xj} - \tilde{s}_{yi}\tilde{s}_{yj}) + C_{-2}^2(\tilde{s}_{xi}\tilde{s}_{yj} + \tilde{s}_{yi}\tilde{s}_{xj}) \Big]$$
$$+ j_2(QR_{ij}) \Big[ C_1^2(\tilde{s}_{zi}\tilde{s}_{xj} + \tilde{s}_{xi}\tilde{s}_{zj}) + C_{-1}^2(\tilde{s}_{zi}\tilde{s}_{yj} + \tilde{s}_{yi}\tilde{s}_{zj}) \Big] \Big\}, \quad (7)$$

where

$$C_0^2 = \tfrac{1}{2}\Big[ 3(\hat{\mathbf{R}}_{ij})_z^2 - 1 \Big]$$
$$C_2^2 = (\hat{\mathbf{R}}_{ij})_x^2 - (\hat{\mathbf{R}}_{ij})_y^2$$
$$C_{-2}^2 = 2(\hat{\mathbf{R}}_{ij})_x(\hat{\mathbf{R}}_{ij})_y \qquad . \quad (8)$$
$$C_1^2 = (\hat{\mathbf{R}}_{ij})_x(\hat{\mathbf{R}}_{ij})_z$$
$$C_{-1}^2 = (\hat{\mathbf{R}}_{ij})_y(\hat{\mathbf{R}}_{ij})_z$$

The unit vector, $\hat{\mathbf{R}}_{ij} \equiv \mathbf{R}_{ij}/|\mathbf{R}_{ij}|$ has Cartesian components $(\hat{\mathbf{R}}_{ij})_\alpha$, $\alpha = x, y, z$. The ordered product $\tilde{s}_{\alpha i}\tilde{s}_{\beta j}$ is defined in Eq. (9),

$$\tilde{s}_{\alpha i}\tilde{s}_{\beta j} \equiv \langle n|\hat{s}_{\alpha i}|m\rangle\langle m|\hat{s}_{\beta j}|n\rangle. \quad (9)$$

Eq. (7) involves spherical Bessel functions $j_0(x)$ and $j_2(x)$,

$$j_0(x) = \frac{\sin(x)}{x}$$
$$j_2(x) = \left(\frac{3}{x^2} - 1\right)\frac{\sin(x)}{x} - \frac{3\cos(x)}{x^2} \quad . \quad (10)$$

For deriving universal **Q**-dependencies, we suppose that a transition takes place between levels that transform according to irreducible representations $\Gamma_n$ and $\Gamma_m$ of the anisotropic point-group, where $\Gamma_n = \Gamma_m$ is permitted. If levels comprise multiple states (for multidimensional $\Gamma_n$ or $\Gamma_m$) intensities are summed over all combinations of initial and final states (that is, we sum over components $k_n$ and $k_m$ of $\Gamma_n$ and $\Gamma_m$, respectively), yielding $L_{\Gamma_n\Gamma_m}(\mathbf{Q})$. Under combined permutations and rotations, the $N$ site-spin operators $\{\hat{\mathbf{s}}_i\}$ span $\Gamma^{(3N)}$. The decomposition of $\Gamma^{(3N)}$ in terms of irreducible representations is formally similar to a symmetry-classification of the translational, rotational and vibrational modes [31] (the difference is that the position vectors $\{\hat{\mathbf{R}}_i\}$ of the magnetic ions are permuted, rotated and inverted by improper rotations, whereas the $\{\hat{\mathbf{s}}_i\}$ are only permuted and rotated).

The derivation of universal **Q**-dependencies proceeds in close analogy to the isotropic case [31]. We assume that $\Gamma^{(3N)}$ contains exactly one irreducible representation $\Gamma_l$ such that $\Gamma_n^* \times \Gamma_l \times \Gamma_m$ contains the totally symmetric $\Gamma_1$, and further assume that $\Gamma_l$ occurs exactly one time in $\Gamma^{(3N)}$. The unitary **v** mediates the transformation to the symmetry-adapted basis,

$$\hat{T}_{\Gamma_q k_q} = \sum_{i\gamma} v_{\Gamma_q k_q, i\gamma} \hat{s}_{i\gamma} \;, \tag{11}$$

where $\Gamma_q k_q$ is a compound index ($\Gamma_q k_q = 1, 2, ..., 3N$) which can include a specific irreducible representation multiple times; $i$ is a site index and $\gamma = x, y, z$. The reverse transformation is given in Eq. (12),

$$\hat{s}_{i\gamma} = \sum_{\Gamma_q k_q} v^*_{\Gamma_q k_q, i\gamma} \hat{T}_{\Gamma_q k_q} \;. \tag{12}$$

Then, as transitions can be mediated only by $\hat{T}_{\Gamma_l k_l}$ transition operators, when working in the the standard convention for (complex unitary) representation matrices [44], we overall obtain Eq. (13) (see Ref. [31] for further details),

$$L_{\Gamma_n \Gamma_m}(\mathbf{Q}) = \langle \Gamma_n \| \hat{\mathbf{T}}_{\Gamma_l} \| \Gamma_m \rangle \langle \Gamma_m \| \hat{\mathbf{T}}_{\Gamma_l} \| \Gamma_n \rangle \sum_{i,j} e^{i\mathbf{Q} \cdot (\mathbf{R}_i - \mathbf{R}_j)} \sum_{\alpha,\beta} \left( \delta_{\alpha\beta} - \frac{Q_\alpha Q_\beta}{Q^2} \right) \sum_{k_l} v^*_{\Gamma_l k_l, i\alpha} v_{\Gamma_l k_l, j\beta} \;, \tag{13}$$

where $\langle \Gamma_n \| \hat{\mathbf{T}}_{\Gamma_l} \| \Gamma_m \rangle$ is a point-group reduced matrix element (RME) [44,45], which is left undefined up to an irrelevant constant factor. (Eq. (13) assumes that $\Gamma_n^* \times \Gamma_m$ contains $\Gamma_l$ exactly once. In the icosahedral group, $\Gamma_n^* \times \Gamma_m$ can contain a specific $\Gamma_l$ twice [44], but universal **Q**-dependencies in the icosahedron, dodecahedron or icosidodecahedron, and most likely also in all larger icosahedral polyhedra, do not occur between two levels that both belong to multidimensional representations, cf. our results for the icosahedron presented below.) In Eq. (13), note the single complex-conjugation sign in $v^*_{\Gamma_l k_l, i\alpha} v_{\Gamma_l k_l, j\beta}$. The **Q**-dependent part of Eq. (13) is $K_{\Gamma_l}(\mathbf{Q})$,

$$K_{\Gamma_l}(\mathbf{Q}) = \sum_{i,j} e^{i\mathbf{Q} \cdot (\mathbf{R}_i - \mathbf{R}_j)} \sum_{\alpha,\beta} \left( \delta_{\alpha\beta} - \frac{Q_\alpha Q_\beta}{Q^2} \right) \sum_{k_l} v^*_{\Gamma_l k_l, i\alpha} v_{\Gamma_l k_l, j\beta} \;. \tag{14}$$

$K_{\Gamma_l}(\mathbf{Q})$ is obviously independent of the spin dynamics, which is contained in the RMEs. With an analytic representation of the $\mathbf{v}_{\Gamma_l k_l}$ vectors available (generated by the usual projection-operator formalism [46,47]), the $K_{\Gamma_l}(\mathbf{Q})$ functions can be tabulated. However, as the polarization factor $\delta_{\alpha\beta} - Q_\alpha Q_\beta / Q^2$ is still present in Eq. (14), the $K_{\Gamma_l}(\mathbf{Q})$ functions take a less compact form than for unpolarized transitions in isotropic systems (such functions were listed in Ref. [31]). One could still tabulate the contributions of different pair types in a local coordinate system. For brevity, we do not pursue such an approach, but instead report the far more compact expressions for powder-averages $\bar{K}_{\Gamma_l}(Q)$. These are obtained by replacing the Cartesian local spin operators in Eq. (7) by their component along $\Gamma_l k_l$ (that is, $\tilde{s}_{\alpha i}\tilde{s}_{\beta j}$ in Eq. (7) is replaced by $v^*_{\Gamma_l k_l, i\alpha} v_{\Gamma_l k_l, j\beta}$), and by summing over $k_l$.

## 3. Results and Discussion

We found that several types of transitions in anisotropic spin rings and small polyhedra are characterized by universal $K_{\Gamma_l}(\mathbf{Q})$ functions. As examples, we chose planar six- and eight-membered rings, as well as the cube and the icosahedron. These systems are illustrated in Figure 1. As we comment below, a certain number of $K_{\Gamma_l}(\mathbf{Q})$ functions exist also in non-planar even-membered spin rings and in other polyhedra (e.g. the tetrahedron, octahedron, cuboctahedron or dodecahedron).

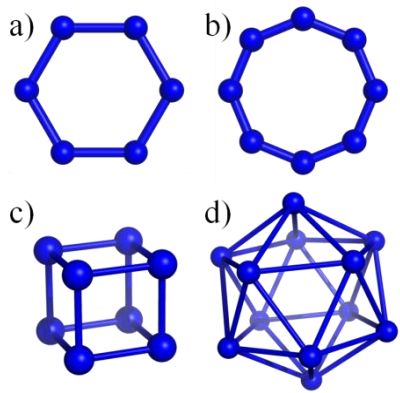

Figure 1: Systems studied in the present work: regular hexagon (a), octagon (b), cube (c), and icosahedron (d). Each sphere marks a spin site. These pictures were generated with the PyMol software.

As explained in the Theory section, a transition type $\Gamma_n \to \Gamma_m$ is associated with a definite $K_{\Gamma_l}(\mathbf{Q})$ if $\Gamma^{(3N)}$ contains exactly one $\Gamma_l$ a single time, such that $\Gamma_n^* \times \Gamma_l \times \Gamma_m$ contains $\Gamma_1$. Thus, only the direct-product table for a given point group [44] and the symmetry-decomposition of $\Gamma^{(3N)}$ are needed to determine if $\Gamma_n \to \Gamma_m$ has a definite $K_{\Gamma_l}(\mathbf{Q})$. Decompositions of $\Gamma^{(3N)}$ are collected in Table 1. For the sake of reference, Table 1 also includes the state-space decomposition for the respective $s = \frac{1}{2}$ systems (but keep in mind that all our results are valid for arbitrary $s$).

Table 1: Decompositions of $\Gamma^{(3N)}$ spanned by $\{\hat{\mathbf{s}}_i\}$ in the anisotropic systems considered in this work.[a] Mulliken symmetry labels follow Ref. [44]. In the last column, the Hilbert-space decompositions for the respective $s = \frac{1}{2}$ systems are provided (for reference only).

| System | $N$ | Group | $\Gamma^{(3N)}$ | $\Gamma(s = \frac{1}{2})$ |
|---|---|---|---|---|
| Ring | 6 | $D_{6h}$ | $A_{1u} \oplus A_{2g} \oplus A_{2u} \oplus B_{1g}$ $\oplus B_{2g} \oplus B_{2u} \oplus 2E_{1g}$ $\oplus E_{1u} \oplus E_{2g} \oplus 2E_{2u}$ | $A_{1g} \oplus 2A_{1u} \oplus 5A_{2g} \oplus 2A_{2u}$ $\oplus 4B_{1g} \oplus 5B_{2g} \oplus 4B_{2g} \oplus B_{2u}$ $\oplus 6E_{1g} \oplus 5E_{1u} \oplus 5E_{2g} \oplus 4E_{2u}$ |
| Ring | 8 | $D_{8h}$ | $A_{1u} \oplus A_{2g} \oplus A_{2u} \oplus B_{1g} \oplus B_{1u}$ $\oplus B_{2u} \oplus 2E_{1g} \oplus E_{1u} \oplus E_{2g}$ $\oplus 2E_{2u} \oplus 2E_{3g} \oplus E_{3u}$ | $13A_{1g} \oplus 8A_{1u} \oplus 5A_{2g} \oplus 8A_{2u}$ $\oplus 14B_{1g} \oplus 8B_{1u} \oplus 6B_{2g} \oplus 8B_{2u}$ $\oplus 16E_{1g} \oplus 14E_{1u} \oplus 17E_{2g} \oplus 16E_{2u}$ $\oplus 16E_{3g} \oplus 14E_{3u}$ |
| Cube | 8 | $O_h$ | $A_{1u} \oplus A_{2g} \oplus E_g \oplus E_u \oplus 2T_{1g}$ $\oplus T_{1u} \oplus T_{2g} \oplus 2T_{2u}$ | $11A_{1g} \oplus 8A_{1u} \oplus 5A_{2g} \oplus 4A_{2u}$ $\oplus 15E_g \oplus 9E_u \oplus 14T_{1g}$ $\oplus 13T_{1u} \oplus 16T_{2g} \oplus 17T_{2u}$ |
| Icosahedron | 12 | $I_h$ | $A_u \oplus 2T_{1g} \oplus T_{1u} \oplus T_{2g} \oplus F_g$ $\oplus F_u \oplus H_g \oplus 2H_u$ | $47A_g \oplus 45A_u \oplus 98T_{1g}$ $\oplus 94T_{1u} \oplus 95T_{2g} \oplus 93T_{2u}$ $\oplus 141F_g \oplus 135F_u$ $\oplus 178H_g \oplus 174H_u$ |

[a]Except for the $N = 8$ ring, the $\Gamma^{(3N)}$ decompositions were already reported in Ref. [31].

The combined spin permutations and global spin rotations that generate the $D_{6h}$ symmetry group were detailed in Ref. [31]. Table 2 lists the $\Gamma_l$ species defining $K_{\Gamma_l}(\mathbf{Q})$ for all $\Gamma_n \to \Gamma_m$ in the $D_{6h}$ ring. Some transitions are not characterized by a universal $\mathbf{Q}$-dependence, because $\Gamma_n$ and $\Gamma_m$ are coupled by multiple transition operators, or transitions have zero intensity, because the set $\{\hat{\mathbf{s}}_i\}$ does not span $A_{1g}$ or $B_{1u}$.

Table 2: Symmetry labels $\Gamma_l$ specifying $K_{\Gamma_l}(\mathbf{Q})$ functions for INS transitions $\Gamma_n \to \Gamma_m$ in the anisotropic $D_{6h}$ spin ring.[a]

|        | $A_{1g}$ | $A_{1u}$ | $A_{2g}$ | $A_{2u}$ | $B_{1g}$ | $B_{1u}$ | $B_{2g}$ | $B_{2u}$ | $E_{1g}$ | $E_{1u}$ | $E_{2g}$ | $E_{2u}$ |
|--------|----------|----------|----------|----------|----------|----------|----------|----------|----------|----------|----------|----------|
| $A_{1g}$ | 0 | $A_{1u}$ | $A_{2g}$ | $A_{2u}$ | $B_{1g}$ | 0 | $B_{2g}$ | $B_{2u}$ | N/A | $E_{1u}$ | $E_{2g}$ | N/A |
| $A_{1u}$ | $A_{1u}$ | 0 | $A_{2u}$ | $A_{2g}$ | 0 | $B_{1g}$ | $B_{2u}$ | $B_{2g}$ | $E_{1u}$ | N/A | N/A | $E_{2g}$ |
| $A_{2g}$ | $A_{2g}$ | $A_{2u}$ | 0 | $A_{1u}$ | $B_{2g}$ | $B_{2u}$ | $B_{1g}$ | 0 | N/A | $E_{1u}$ | $E_{2g}$ | N/A |
| $A_{2u}$ | $A_{2u}$ | $A_{2g}$ | $A_{1u}$ | 0 | $B_{2u}$ | $B_{2g}$ | 0 | $B_{1g}$ | $E_{1u}$ | N/A | N/A | $E_{2g}$ |
| $B_{1g}$ | $B_{1g}$ | 0 | $B_{2g}$ | $B_{2u}$ | 0 | $A_{1u}$ | $A_{2g}$ | $A_{2u}$ | $E_{2g}$ | N/A | N/A | $E_{1u}$ |
| $B_{1u}$ | 0 | $B_{1g}$ | $B_{2u}$ | $B_{2g}$ | $A_{1u}$ | 0 | $A_{2u}$ | $A_{2g}$ | N/A | $E_{2g}$ | $E_{1u}$ | N/A |
| $B_{2g}$ | $B_{2g}$ | $B_{2u}$ | $B_{1g}$ | 0 | $A_{2g}$ | $A_{2u}$ | 0 | $A_{1u}$ | $E_{2g}$ | N/A | N/A | $E_{1u}$ |
| $B_{2u}$ | $B_{2u}$ | $B_{2g}$ | 0 | $B_{1g}$ | $A_{2u}$ | $A_{2g}$ | $A_{1u}$ | 0 | N/A | $E_{2g}$ | $E_{1u}$ | N/A |
| $E_{1g}$ | N/A | $E_{1u}$ | N/A | $E_{1u}$ | $E_{2g}$ | N/A | $E_{2g}$ | N/A | N/A | N/A | N/A | N/A |
| $E_{1u}$ | $E_{1u}$ | N/A | $E_{1u}$ | N/A | N/A | $E_{2g}$ | N/A | $E_{2g}$ | N/A | N/A | N/A | N/A |
| $E_{2g}$ | $E_{2g}$ | N/A | $E_{2g}$ | N/A | N/A | $E_{1u}$ | N/A | $E_{1u}$ | N/A | N/A | N/A | N/A |
| $E_{2u}$ | N/A | $E_{2g}$ | N/A | $E_{2g}$ | $E_{1u}$ | N/A | $E_{1u}$ | N/A | N/A | N/A | N/A | N/A |

[a]The point-group labels $\Gamma_n$ and $\Gamma_m$ of the two levels are given in boldface in the first row and first column. The $\Gamma_l$ are given in the bulk of the table. Forbidden transitions have entry 0, whereas N/A marks transitions which do not have a universal **Q**-dependence. The table is symmetric about the diagonal.

As remarked, the single-crystal $K_{\Gamma_l}(\mathbf{Q})$ functions cannot be given in a particularly compact form and we instead tabulate powder averages $\bar{K}_{\Gamma_l}(Q)$. Each $\bar{K}_{\Gamma_l}(Q)$ is an expansion in $j_0(Qr_t)$ and $j_2(Qr_t)$,

$$\bar{K}_{\Gamma_l}(Q) = \sum_t \left[ c_t^{\Gamma_l} j_0(Qr_t) + d_t^{\Gamma_l} j_2(Qr_t) \right], \quad (15)$$

where the sum runs over all pairs $t$ that are not equivalent by symmetry, and $r_t$ is the Cartesian distance between two ions forming a pair. The expansion coefficients $c_t^{\Gamma_l}$ and $d_t^{\Gamma_l}$ (in the following, these are simply denoted by $c$ and $d$, respectively) and one representative pair defining each group $t$ are collected in Table 3. We formally count (1, 1) as a pair, with $r_t = 0$ and $d_t^{\Gamma_l} = 0$ (the vanishing $d$ coefficients for this pair are not explicitly included in Table 3 or any of the following tables).

Table 3: Coefficients $c$ and $d$ ($c_t^{\Gamma_l}$ and $d_t^{\Gamma_l}$ in Eq. (15)) defining $\bar{K}_{\Gamma_l}(Q)$ functions for the anisotropic $D_{6h}$ spin ring. Sites are numbered consecutively around the ring.[a]

|        |   | $A_{1u}$ | $A_{2g}$ | $A_{2u}$ | $B_{1g}$ | $B_{2g}$ | $B_{2u}$ | $E_{1u}$ | $E_{2g}$ |
|--------|---|------|------|------|------|------|------|------|------|
| (1, 1) | $c$ | 4 | 2 | 4 | 4 | 4 | 2 | 2 | 2 |
| (1, 2) | $c$ | 4 | 4 | 4 | -4 | -4 | -4 | 2 | -2 |
|        | $d$ | -5 | -2 | 7 | 5 | -7 | 2 | -1 | 1 |
| (1, 3) | $c$ | -4 | 4 | -4 | -4 | -4 | 4 | -2 | -2 |
|        | $d$ | -7 | -2 | 5 | -7 | 5 | -2 | 1 | 1 |
| (1, 4) | $c$ | -4 | 2 | -4 | 4 | 4 | -2 | -2 | 2 |
|        | $d$ | -4 | -1 | 2 | 4 | -2 | 1 | 1 | -1 |

[a]One representative site pair $(i, j)$ for each group $t$ of symmetry-equivalent pairs is given in the first column. The $\Gamma_l$ label is specified in the first row.

As a concrete example for constructing $\bar{K}_{\Gamma_l}(Q)$ functions from Table 3, we explicitly write out $\bar{K}_{A_{1u}}(Q)$ in Eq. (16),

$$\bar{K}_{A_{1u}}(Q) = 4 + 4j_0(QR_{12}) - 5j_2(QR_{12}) \\ - 4j_0(QR_{13}) - 7j_2(QR_{13}) - 4j_0(QR_{14}) - 4j_2(QR_{14}) \quad . \tag{16}$$

The coefficients in Table 3 (and following tables) are provided in the simplest possible form. That is, each result from an evaluation of Eq. (7) (which is here implied to include a summation over all components of multidimensional representations, see Theory section) was multiplied by a specific factor to avoid fractions in $c$ and $d$. Consequently, these coefficient sets do not in general reflect the relative intensities for transitions that originate in first order from the same spin multiplet (when sublevels merge into a single spin multiplet, the $d$ contributions must vanish, so that $\bar{K}(Q)$ becomes an expansion in $j_0(Qr_t)$ only). It is a straightforward exercise in group theory to determine anisotropy-induced symmetry-decompositions of spin multiplets with definite SPS (physically speaking, this could be loosely regarded as a combined crystal-field and spin-orbit splitting), but this issue is of no immediate concern here; a few examples can be found in Refs. [39,48,49].

The sum of $c$ coefficients, $\sum_t c_t^{\Gamma_l}$, is non-zero for $\bar{K}_{A_{2g}}(Q)$ only, where the $A_{2g}$ transition operator is $\hat{S}_z$ and $\bar{K}_{A_{2g}}(Q)$ has its global maximum at $Q = 0$. All the other $\bar{K}_{\Gamma_l}(Q)$ vanish at $Q = 0$. The universal $\bar{K}_{\Gamma_l}(Q)$ functions for the $D_{6h}$ spin ring are plotted in Figure 2. In Figure 2 we also illustrate the transition operators in terms of the sets of $N$ three-dimensional vectors defined by $\mathbf{v}_{\Gamma_l k_l}$ (cf. Eq. (11)). To thus draw the transition operators as vectors attached to individual spin sites, the complex components of $\mathbf{v}_{\Gamma_l k_l}$ were eliminated through appropriate unitary transformations on the components of multi-dimensional representations ($E_{1u}$ or $E_{2g}$ in the present example).

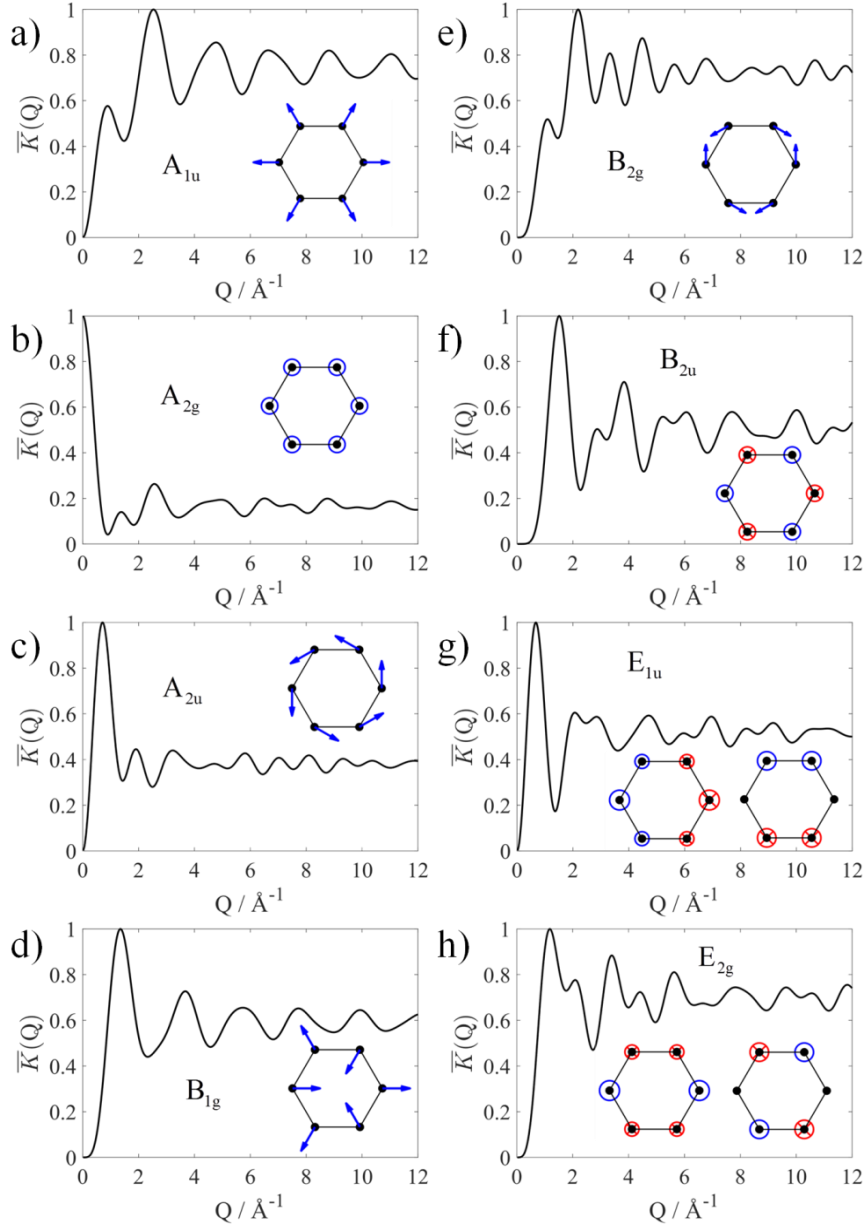

Figure 2: Universal powder-averaged functions $\bar{K}_{\Gamma_l}(Q)$ (in arbitrary units, with maximal intensities normalized to 1; symmetry labels $\Gamma_l$ are specified in the plots) in the anisotropic $D_{6h}$ spin ring. Magnetic ions have a nearest-neighbor distance of 3Å and lie in the $xy$-plane. Transition operators are drawn in terms of vectors attached to the spin centers. They are either confined to the plane (panels a–e) or parallel to the $z$-axis (f–h). Vectors pointing in the positive or negative $z$-direction are shown as blue circles, or red circles with crosses, respectively (f–h). For $E_{1u}$ and $E_{2g}$ (g and h), vectors marked by large circles have twice the length of vectors marked by small circles.

While the curves shown in Figure 2 were computed by evaluating the analytical representations of the $\bar{K}_{\Gamma_l}(Q)$ functions, we checked that the same curves are obtained in a numerical evaluation of the spherical integral (Eq. (6); Lebedev-Laikov grids [50] were used) based on exact eigenstates of the $s=\tfrac{1}{2}$ spin Hamiltonian with nearest-neighbor isotropic and symmetric anisotropic interactions. The basis was adapted to $D_{6h}$ symmetry and exact energy spectra were compared against results from calculations that did not employ symmetry. Results for the other systems discussed below were similarly verified.

Turning to the $D_{8h}$ ring, the $\Gamma_l$ species are collected as a function of $\Gamma_n \to \Gamma_m$ in Table 4 and the $c$ and $d$ coefficients defining $\bar{K}_{\Gamma_l}(Q)$ functions are given in Table 5. $\bar{K}_{\Gamma_l}(Q)$ plots and the corresponding transition operators are finally shown in Figure 3.

Table 4: Symmetry labels $\Gamma_l$ specifying $K_{\Gamma_l}(\mathbf{Q})$ functions for INS transitions $\Gamma_n \to \Gamma_m$ in the anisotropic $D_{8h}$ spin ring. For further details, see footnote to Table 2.

|     | $A_{1g}$ | $A_{1u}$ | $A_{2g}$ | $A_{2u}$ | $B_{1g}$ | $B_{1u}$ | $B_{2g}$ | $B_{2u}$ | $E_{1g}$ | $E_{1u}$ | $E_{2g}$ | $E_{2u}$ | $E_{3g}$ | $E_{3u}$ |
|---|---|---|---|---|---|---|---|---|---|---|---|---|---|---|
| $A_{1g}$ | 0 | $A_{1u}$ | $A_{2g}$ | $A_{2u}$ | $B_{1g}$ | $B_{1u}$ | 0 | $B_{2u}$ | N/A | $E_{1u}$ | $E_{2g}$ | N/A | N/A | $E_{3u}$ |
| $A_{1u}$ | $A_{1u}$ | 0 | $A_{2u}$ | $A_{2g}$ | $B_{1u}$ | $B_{1g}$ | $B_{2u}$ | 0 | $E_{1u}$ | N/A | N/A | $E_{2g}$ | $E_{3u}$ | N/A |
| $A_{2g}$ | $A_{2g}$ | $A_{2u}$ | 0 | $A_{1u}$ | 0 | $B_{2u}$ | $B_{1g}$ | $B_{1u}$ | N/A | $E_{1u}$ | $E_{2g}$ | N/A | N/A | $E_{3u}$ |
| $A_{2u}$ | $A_{2u}$ | $A_{2g}$ | $A_{1u}$ | 0 | $B_{2u}$ | 0 | $B_{1u}$ | $B_{1g}$ | $E_{1u}$ | N/A | N/A | $E_{2g}$ | $E_{3u}$ | N/A |
| $B_{1g}$ | $B_{1g}$ | $B_{1u}$ | 0 | $B_{2u}$ | 0 | $A_{1u}$ | $A_{2g}$ | $A_{2u}$ | N/A | $E_{3u}$ | $E_{2g}$ | N/A | N/A | $E_{1u}$ |
| $B_{1u}$ | $B_{1u}$ | $B_{1g}$ | $B_{2u}$ | 0 | $A_{1u}$ | 0 | $A_{2u}$ | $A_{2g}$ | $E_{3u}$ | N/A | N/A | $E_{2g}$ | $E_{1u}$ | N/A |
| $B_{2g}$ | 0 | $B_{2u}$ | $B_{1g}$ | $B_{1u}$ | $A_{2g}$ | $A_{2u}$ | 0 | $A_{1u}$ | N/A | $E_{3u}$ | $E_{2g}$ | N/A | N/A | $E_{1u}$ |
| $B_{2u}$ | $B_{2u}$ | 0 | $B_{1u}$ | $B_{1g}$ | $A_{2u}$ | $A_{2g}$ | $A_{1u}$ | 0 | $E_{3u}$ | N/A | N/A | $E_{2g}$ | $E_{1u}$ | N/A |
| $E_{1g}$ | N/A | $E_{1u}$ | N/A | $E_{1u}$ | N/A | $E_{3u}$ | N/A | $E_{3u}$ | N/A | N/A | N/A | N/A | N/A | N/A |
| $E_{1u}$ | $E_{1u}$ | N/A | $E_{1u}$ | N/A | $E_{3u}$ | N/A | $E_{3u}$ | N/A | N/A | N/A | N/A | N/A | N/A | N/A |
| $E_{2g}$ | $E_{2g}$ | N/A | $E_{2g}$ | N/A | $E_{2g}$ | N/A | $E_{2g}$ | N/A | N/A | N/A | N/A | N/A | N/A | N/A |
| $E_{2u}$ | N/A | $E_{2g}$ | N/A | $E_{2g}$ | N/A | $E_{2g}$ | N/A | $E_{2g}$ | N/A | N/A | N/A | N/A | N/A | N/A |
| $E_{3g}$ | N/A | $E_{3u}$ | N/A | $E_{3u}$ | N/A | $E_{1u}$ | N/A | $E_{1u}$ | N/A | N/A | N/A | N/A | N/A | N/A |
| $E_{3u}$ | $E_{3u}$ | N/A | $E_{3u}$ | N/A | $E_{1u}$ | N/A | $E_{1u}$ | N/A | N/A | N/A | N/A | N/A | N/A | N/A |

Table 5: Coefficients $c$ and $d$ defining $\bar{K}_{\Gamma_l}(Q)$ functions for the anisotropic $D_{8h}$ spin ring. For further details, see footnote to Table 3.

| | | $A_{1u}$ | $A_{2g}$ | $A_{2u}$ | $B_{1g}$ | $B_{1u}$ | $B_{2u}$ | $E_{1u}$ | $E_{2g}$ | $E_{3u}$ |
|---|---|---|---|---|---|---|---|---|---|---|
| (1, 1) | $c$ | 4 | 2 | 4 | 2 | 4 | 4 | 4 | 2 | 4 |
| (1, 2) | $c$ | $4\sqrt{2}$ | 4 | $4\sqrt{2}$ | -4 | $-4\sqrt{2}$ | $-4\sqrt{2}$ | $4\sqrt{2}$ | 0 | $-4\sqrt{2}$ |
| | $d$ | $\sqrt{2}-6$ | -2 | $\sqrt{2}+6$ | 2 | $-\sqrt{2}-6$ | $-\sqrt{2}+6$ | $-2\sqrt{2}$ | 0 | $2\sqrt{2}$ |
| (1, 3) | $c$ | 0 | 4 | 0 | 4 | 0 | 0 | 0 | -4 | 0 |
| | $d$ | -6 | -2 | 6 | -2 | 6 | -6 | 0 | 2 | 0 |
| (1, 4) | $c$ | $-4\sqrt{2}$ | 4 | $-4\sqrt{2}$ | -4 | $4\sqrt{2}$ | $4\sqrt{2}$ | $-4\sqrt{2}$ | 0 | $4\sqrt{2}$ |
| | $d$ | $-\sqrt{2}-6$ | -2 | $-\sqrt{2}+6$ | 2 | $\sqrt{2}-6$ | $\sqrt{2}+6$ | $2\sqrt{2}$ | 0 | $-2\sqrt{2}$ |
| (1, 5) | $c$ | -4 | 2 | -4 | 2 | -4 | -4 | -4 | 2 | -4 |
| | $d$ | -4 | -1 | 2 | -1 | 2 | -4 | 2 | -1 | 2 |

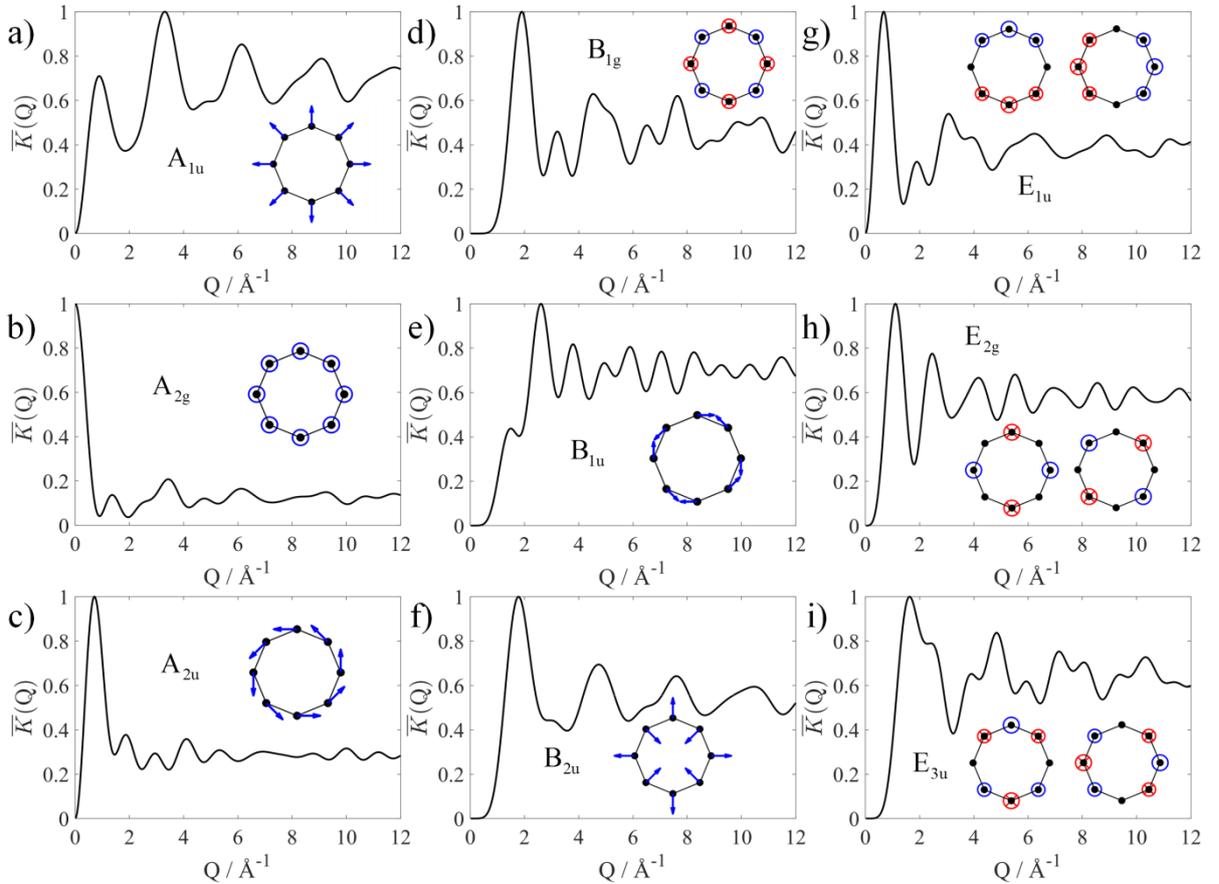

Figure 3: Universal $\bar{K}_{\Gamma_l}(Q)$ functions and transition operators in the anisotropic $D_{8h}$ spin ring. For further details, see caption to Figure 2. For $E_{1u}$ and $E_{3u}$ (panels g and i), vectors marked by large circles have a length which is larger by a factor of $\sqrt{2}$ compared to vectors marked by small circles.

The basic features observed in the $D_{6h}$ and $D_{8h}$ systems are generally valid in $D_{Nh}$ for larger $N$ (with $N$ even). That is, transitions between two non-degenerate levels are either forbidden or have a universal $K_{\Gamma_l}(\mathbf{Q})$, because each one-dimensional representation occurs at most one time in $\Gamma^{(3N)}$. Although a remarkable variety of symmetric spin rings (particularly those with even $N$) were already studied by INS (a non-exhaustive list for $N = 8, 10, 18$ includes Refs. [21,51–58]), to the best of our knowledge, none of these molecules had $D_{Nh}$ symmetry. One could then ask, if any universal $\mathbf{Q}$-dependencies still occur in the more common group $D_N$ (which lacks the inversion center). Indeed, in $D_6$, $A_1$ and $B_1$ occur only once in $\Gamma^{(3N)}$ and in $D_8$ this applies to $A_1$ and $B_2$, with analogous relations for larger $N$ (with $N$ even). Thus, two distinct universal $\mathbf{Q}$-dependent functions still occur in anisotropic $D_N$ spin rings. The $\bar{K}_{\Gamma_l}(Q)$ functions are the same as in the $D_{Nh}$ systems (assuming that the nearest-neighbor distance is the same). That is, $\bar{K}_{A_1}(Q)$ and $\bar{K}_{B_1}(Q)$ in the $D_6$ ring are the same as $\bar{K}_{A_{1u}}(Q)$ and $\bar{K}_{B_{1g}}(Q)$ in $D_{6h}$, respectively, etc.

As a brief comment on odd-membered spin rings, we consider the $D_{5h}$ system, where

$$\Gamma^{(15)} = A_1'' \oplus A_2' \oplus A_2'' \oplus E_1' \oplus 2E_1'' \oplus E_2' \oplus 2E_2'' \ . \tag{17}$$

Eigenstates of the spin Hamiltonian transform as fermionic representations ($E_{1/2}, E_{3/2}, E_{5/2}, E_{7/2}$, and $E_{9/2}$, in the notation of Ref. [44]) of the double-group $D_{5h}^*$ [38]. The direct-product table for such representations [44] shows that there is no combination $\Gamma_n \to \Gamma_m$ such that the above-given conditions for a universal $\mathbf{Q}$-dependence are fulfilled. This appears to be the case for all odd-membered spin rings, but no proof shall be attempted here.

We now turn to the spin cube. A number of cubic spin clusters has been synthesized, although apparently no system with genuine cubic symmetry was realized yet. We here consider a perfect cube ($O_h$ group). Only transitions involving at least one one-dimensional representation are explicitly included in Table 6, because only these can have a universal $K_{\Gamma_l}(\mathbf{Q})$ in the cube.

Table 6: $O_h$ symmetry labels $\Gamma_l$ defining $K_{\Gamma_l}(\mathbf{Q})$ functions for INS transitions $\Gamma_n \rightarrow \Gamma_m$ in the anisotropic spin cube. For further explanations, see the footnote to Table 2.[a]

|          | $A_{1g}$ | $A_{2g}$ | $E_g$ | $T_{1g}$ | $T_{2g}$ | $A_{1u}$ | $A_{2u}$ | $E_u$ | $T_{1u}$ | $T_{2u}$ |
|----------|----------|----------|-------|----------|----------|----------|----------|-------|----------|----------|
| $A_{1g}$ | 0        | $A_{2g}$ | $E_g$ | N/A      | $T_{2g}$ | $A_{1u}$ | 0        | $E_u$ | $T_{1u}$ | N/A      |
| $A_{2g}$ | $A_{2g}$ | 0        | $E_g$ | $T_{2g}$ | N/A      | 0        | $A_{1u}$ | $E_u$ | N/A      | $T_{1u}$ |
| $A_{1u}$ | $A_{1u}$ | 0        | $E_u$ | $T_{1u}$ | N/A      | 0        | $A_{2g}$ | $E_g$ | N/A      | $T_{2g}$ |
| $A_{2u}$ | 0        | $A_{1u}$ | $E_u$ | N/A      | $T_{1u}$ | $A_{2g}$ | 0        | $E_g$ | $T_{2g}$ | N/A      |

[a] Only pairs with at least one one-dimensional level ($A_{1g}$, $A_{2g}$, $A_{1u}$ or $A_{2u}$) are included, because all the other entries would be N/A.

Table 7: Coefficients $c$ and $d$ defining $\bar{K}_{\Gamma_l}(Q)$ functions for the anisotropic $O_h$ spin cube. For further details, see footnote to Table 3.

|        |   | $A_{1u}$ | $A_{2g}$ | $E_g$ | $E_u$ | $T_{1u}$ | $T_{2g}$ |
|--------|---|----------|----------|-------|-------|----------|----------|
| (1, 1) | $c$ | 2 | 2 | 4 | 4 | 4 | 4 |
| (1, 2) | $c$ | 2 | -2 | -4 | 4 | 4 | -4 |
|        | $d$ | -4 | 4 | 8 | -8 | 4 | -4 |
| (1, 3) | $c$ | -2 | -2 | -4 | -4 | -4 | -4 |
|        | $d$ | -5 | -5 | -1 | -1 | 5 | 5 |
| (1, 4) | $c$ | -2 | 2 | 4 | -4 | -4 | 4 |
|        | $d$ | -2 | 2 | -2 | 2 | 2 | -2 |

The centers defining distinct pair-types in Table 7 are numbered in the order of increasing distance from site 1. For convenience, the site numbering is also given in a planar coupling graph in Figure 4a. (In general, two pairs which are not symmetry-equivalent can still have the same Cartesian distance, but this caveat is of no concern for the cube or icosahedron.)

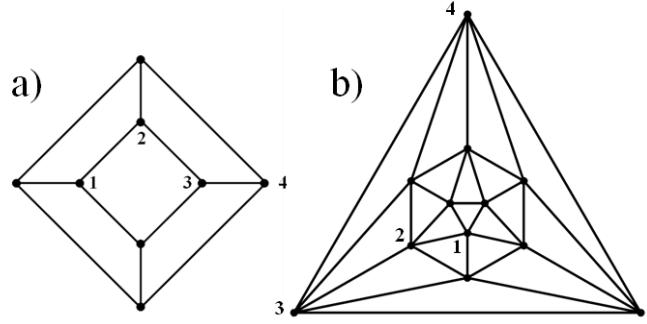

Figure 4: Schlegel diagrams for the cube (a) and the icosahedron (b), including a numbering of sites forming distinct pairs with site 1.

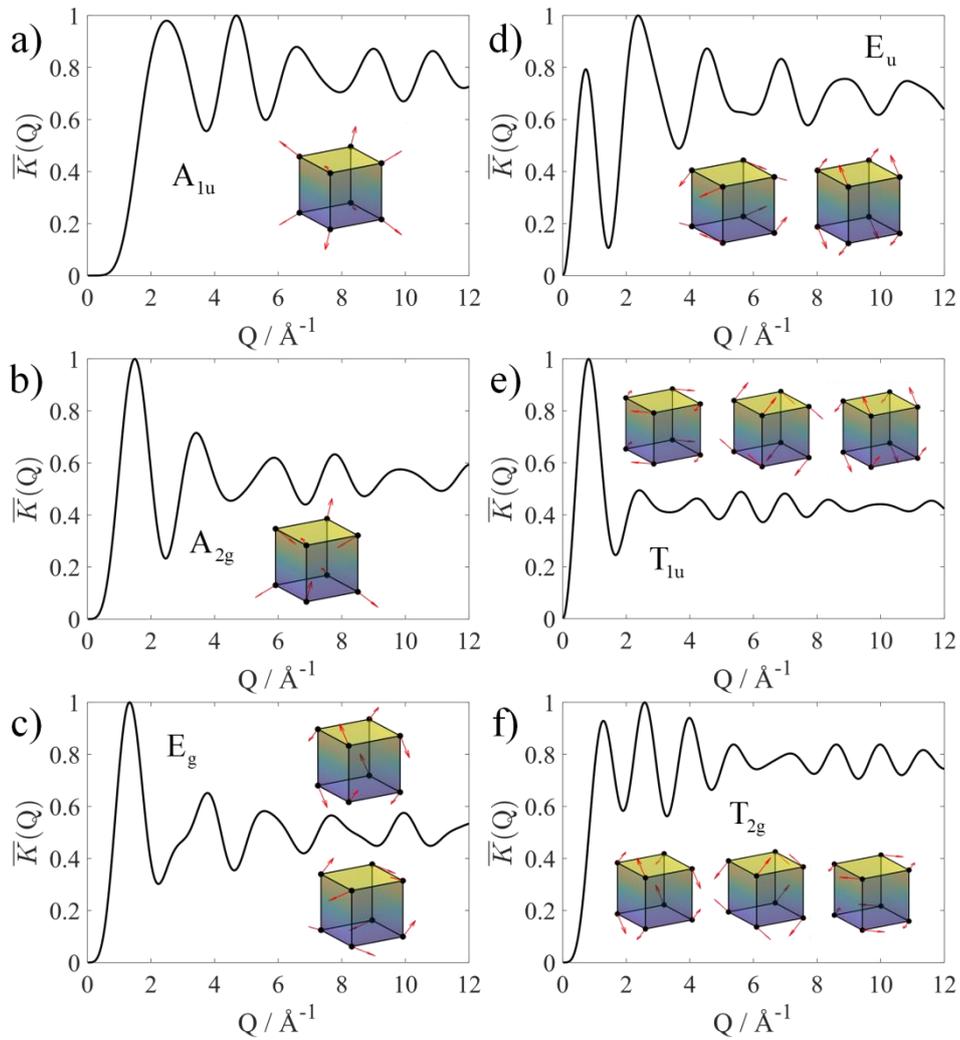

Figure 5: Universal $\bar{K}_{\Gamma_l}(Q)$ functions and transition operators in the anisotropic spin cube. For further details, see caption to Figure 2. In the $A_{1u}$ transition operator, all vectors are directed radially outward, while the vectors alternately point inward and outward in the $A_{2g}$ operator. The multidimensional transition operators are slightly more complicated.

The $\bar{K}_{\Gamma_l}(Q)$ functions and transition operators are plotted in Figure 5. With regard to checking these curves in numerical calculations based on exact eigenstates, it should be mentioned that the anisotropic cube with symmetric exchange between nearest neighbors has an artificially high symmetry [31], which causes accidental degeneracies. This additional symmetry is lifted by symmetric anisotropic exchange between next-nearest neighbors, which yields a spin model with actual $O_h$ symmetry.

The discovery of interesting properties of the classical- and quantum-spin icosahedron [59,60] motivated efforts to chemically synthesize such a species, but so far these efforts yielded only an $\{Fe_9\}$ molecule composed of $Fe^{3+}$ ($s=\frac{5}{2}$) ions [61]. This tridiminished icosahedron was studied by INS [62,63]. If genuine spin icosahedra should become available in the future, INS would represent a method of choice for their detailed magnetic characterization. In the anisotropic icosahedron, most transitions involving at least one non-degenerate level have a definite $K_{\Gamma_l}(\mathbf{Q})$, see Table 8, but no definite $K_{\Gamma_l}(\mathbf{Q})$ exists when $\Gamma_n$ and $\Gamma_m$ are both multi-dimensional.

Table 8: $I_h$ symmetry labels $\Gamma_l$ specifying $K_{\Gamma_l}(\mathbf{Q})$ functions for INS transitions $\Gamma_n \to \Gamma_m$ in the anisotropic spin icosahedron. For further explanations, see the footnote to Table 2.[a]

|     | $A_g$ | $T_{1g}$ | $T_{2g}$ | $F_g$ | $H_g$ | $A_u$ | $T_{1u}$ | $T_{2u}$ | $F_u$ | $H_u$ |
|-----|-------|----------|----------|-------|-------|-------|----------|----------|-------|-------|
| $A_g$ | 0 | N/A | $T_{2g}$ | $F_g$ | $H_g$ | $A_u$ | $T_{1u}$ | 0 | $F_u$ | N/A |
| $A_u$ | $A_u$ | $T_{1u}$ | 0 | $F_u$ | N/A | 0 | N/A | $T_{2g}$ | $F_g$ | $H_g$ |

[a]Only pairs with at least one one-dimensional level ($A_g$ or $A_u$) are included, because all the other entries would be N/A.

Table 9: Coefficients $c$ and $d$ defining $\bar{K}_{\Gamma_l}(Q)$ functions for the anisotropic $I_h$ icosahedron. Sites are numbered by increasing distance from site 1 (cf. Figure 4b). For further details, see footnote to Table 3.

|        |   | $A_u$ | $T_{1u}$ | $T_{2g}$ | $F_g$ | $F_u$ | $H_g$ |
|--------|---|-------|----------|----------|-------|-------|-------|
| (1, 1) | $c$ | 4 | 8 | 4 | 4 | 2 | 8 |
| (1, 2) | $c$ | $4\sqrt{5}$ | $8\sqrt{5}$ | -4 | -4 | $-2\sqrt{5}$ | -8 |
|        | $d$ | $-15+\sqrt{5}$ | $15-\sqrt{5}$ | $-1+3\sqrt{5}$ | $-1-3\sqrt{5}$ | $\sqrt{5}$ | $7+3\sqrt{5}$ |
| (1, 3) | $c$ | $-4\sqrt{5}$ | $-8\sqrt{5}$ | -4 | -4 | $2\sqrt{5}$ | -8 |
|        | $d$ | $-15-\sqrt{5}$ | $15+\sqrt{5}$ | $-1-3\sqrt{5}$ | $-1+3\sqrt{5}$ | $-\sqrt{5}$ | $7-3\sqrt{5}$ |
| (1, 4) | $c$ | -4 | -8 | 4 | 4 | -2 | 8 |
|        | $d$ | -4 | 4 | 4 | -2 | 1 | -4 |

The $\bar{K}_{\Gamma_l}(Q)$ functions are plotted in Figure 6. The curves for the different transition-types take very distinct shapes and should be already experimentally distinguishable by the location of the global maximum on the $Q$-axis.

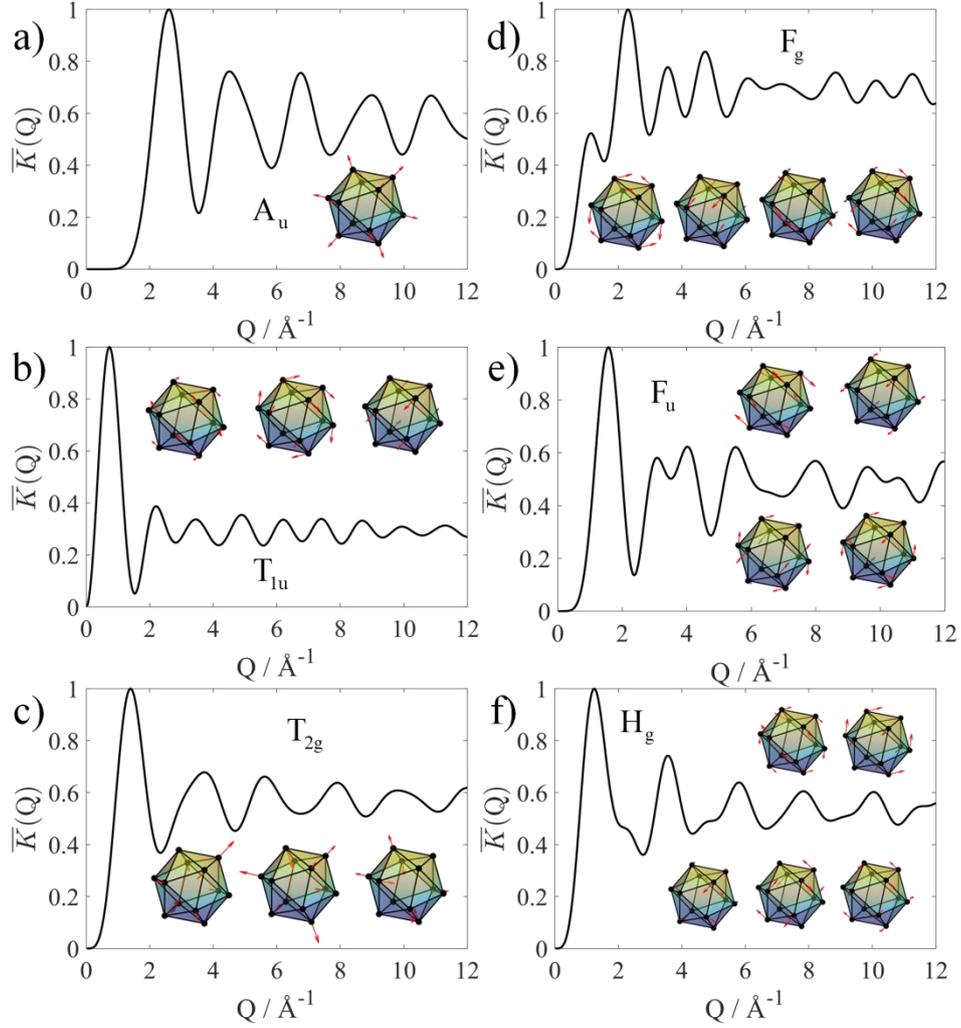

Figure 6: Universal $\bar{K}_{\Gamma_l}(Q)$ functions and transition operators in the anisotropic spin icosahedron. For further details, see caption to Figure 2.

It is lastly worth mentioning that universal $K_{\Gamma_l}(\mathbf{Q})$ functions occur also in many other polyhedra. When there is a $\Gamma_l$ that occurs exactly one time in $\Gamma^{(3N)}$, then transitions mediated by the $\Gamma_l$ transition operator (e.g., $\Gamma_1 \to \Gamma_l$) are characterized by $K_{\Gamma_l}(\mathbf{Q})$. We have explicitly checked that $K_{\Gamma_l}(\mathbf{Q})$ functions exist in the tetrahedron, octahedron, truncated tetrahedron, cuboctahedron, dodecahedron, icosidodecahedron and truncated icosahedron (and presumably also in a number of other polyhedra). However, the number of distinct $K_{\Gamma_l}(\mathbf{Q})$ functions tends to decrease with system size. There are six such functions in the icosahedron (see above), three in the dodecahedron ($K_{A_u}(\mathbf{Q})$, $K_{T_{1u}}(\mathbf{Q})$ and $K_{T_{2u}}(\mathbf{Q})$), but only one in the icosidodecahedron and truncated icosahedron ($K_{A_u}(\mathbf{Q})$ and $K_{A_g}(\mathbf{Q})$, respectively). A more extensive discussion for other systems is beyond the scope of this work.

## 4. Conclusions

It was recently shown [31] that the point-group symmetry of molecular spin clusters can impose selection rules on INS transitions in both isotropic and anisotropic spin models. Besides, the momentum-transfer dependence in isotropic systems is under certain conditions completely determined by the spin-permutational symmetry of the levels [31,36]. In the present work, we have addressed the question if definite universal **Q**-dependencies exist also in anisotropic spin systems. Indeed, dihedral spin rings and spin polyhedra are systems that exhibit transitions that contain dynamical information only in terms of the energy transfer and the transition strength, while the **Q**-dependence exclusively reflects the relative symmetry of the two levels. We have described a simple formalism to generally identify such transitions and applied it to two spin rings ($N = 6, 8$) and two polyhedra (cube and icosahedron). Universal powder-averaged scattering intensities were compactly tabulated. For an efficient accumulation and theoretical modeling of data, it appears advantageous to be aware of those cases where dynamical and geometrical effects are strictly independent. In view of recent significant progress in INS techniques, we therefore believe that the present results could help in future analyses of spectra of symmetric spin clusters.

**Acknowledgements.** The author thanks Martin Kaupp for support.